\documentclass[11pt,titlepage]{article}

\usepackage{setspace} 
\newtheorem{lemma}{Lemma} 

\newtheorem{definition}[lemma]{Definition}
\newtheorem{example}[lemma]{Example}

\usepackage{amsmath}

\usepackage{graphicx}

\usepackage[round,numbers,sort&compress]{natbib} 
\title{The effect of negative feedback loops on the dynamics of Boolean networks}

\author{Eduardo Sontag\thanks{
Corresponding author.  Address: Mathematics Department,
    Rutgers University,
    Piscataway, NJ 08854, USA,
    Tel:~(732)445-3072}\\
Mathematics Department,\\
Rutgers University,    Piscataway, NJ
\and Alan~Veliz-Cuba\\
Virginia Bioinformatics Institute/Department of Mathematics\\
Virginia Polytechnic Institute and State University, Blacksburg, VA
\and Reinhard Laubenbacher\\
Virginia Bioinformatics Institute/Department of Mathematics\\
Virginia Polytechnic Institute and State University, Blacksburg, VA
\and Abdul Salam Jarrah\\
Virginia Bioinformatics Institute/Department of Mathematics\\
Virginia Polytechnic Institute and State University, Blacksburg, VA
}

\date{}

\begin{document}

\maketitle

\abstract{
   Feedback loops play an important role in
   determining the dynamics of biological networks.  In order to study the role
   of negative feedback loops, this paper introduces the notion of
   ``distance to positive feedback (PF-distance)" which in essence captures the
   number of ``independent" negative feedback loops in the
   network, a property inherent in the network topology.
   Through a computational study using Boolean networks it is shown that PF-distance
   has a strong influence on network dynamics and correlates very well with
   the number and length of limit cycles in the phase space of the network.
   To be precise, it is shown that,
   as the number of independent negative feedback loops increases,
   the number (length) of limit cycles tends to decrease (increase).
   These conclusions are consistent with the fact that certain natural
   biological networks exhibit generally
   regular behavior and have fewer negative feedback
   loops than randomized networks with the same numbers of nodes and
   connectivity.

}

\clearpage

\section*{Introduction}


\renewcommand{\cite}[1]{\citet{#1}}

{A}n understanding of the design principles of biochemical networks,
such as gene regulatory, metabolic, or intracellular signaling
networks is a central concern of systems biology.  In particular,
the intricate interplay between network topology and resulting
dynamics is crucial to our understanding of such networks, as is
their presumed modular structure. Features that relate network
topology to dynamics may be considered ``robust'' in the sense that
their influence does not depend on detailed quantitative features
such as exact flux rates. A topological feature of central
interest in this context is the existence of positive and negative
feedback loops. There is broad consensus that feedback loops have
a decisive effect on dynamics, which has been studied extensively
through the analysis of mathematical network models, both
continuous and discrete.
Indeed, it has long been appreciated by biologists that positive
and negative feedback loops play a central role in controlling the
dynamics of a wide range of biological systems. Thomas et al.
\cite{ThoThiKau} conjectured that positive feedback loops are
necessary for multistationarity whereas negative feedback loops
are necessary for the existence of periodic behaviors. Proofs for
different partial cases of these conjectures have been given, see,
e.g., \cite{Soule, Plahte, Gouze, CinDem, AHS}.  Moreover, it is
widely believed \cite{ejc05es} that an abundance of negative loops
should result in the existence of ``chaotic'' behavior in the
network.  This paper provides strong evidence in support of this
latter conjecture.

We focus here on Boolean network (BN) models, a popular model type
for biochemical networks, initially introduced by S. Kauffman
\cite{kauff69}.  In particular, we study BN models in which each
directed edge can be characterized as either an inhibition or an
activation.
In Boolean models of biological networks, each variable can only
attain two values ($0/1$ or ``on/off'').  These values represent
whether a gene is being expressed, or the concentration of a
protein is above a certain threshold, at time $t$. When detailed
information on kinetic rates of protein-DNA or protein-protein
interactions is lacking, and especially if regulatory
relationships are strongly sigmoidal, such models are useful in
theoretical analysis, because they serve to focus attention on the
basic dynamical characteristics while ignoring specifics of
reaction mechanisms, see
\cite{kauffman69b,kauffman-glass73,albertothmer03,chaves_albert_sontagJTB05}).

Boolean networks constructed from monotone Boolean functions
(i.e. each node or ``gate'' computes a function which is increasing on all
arguments)
are of particular interest, and have been studied extensively, in the
electronic circuit design and pattern recognition literature
\cite{Gilbert54,Minsky67}, as well as in the computer science literature;
see e.g. \cite{aracena-1,aracena-2,golesHer}
for recent references.
For Boolean and all other finite iterated systems, all
trajectories must either settle into equilibria or to periodic
orbits, whether the system is made up of monotone functions or
not, but monotone networks have always somewhat shorter cycles.
This is because periodic orbits must be anti-chains, i.e.\ no two
different states can be compared; see~\cite{Gilbert54,smith}. An
upper bound may be obtained by appealing to Sperner's Theorem
(\cite{andersonsperner}): Boolean systems on $n$ variables can
have orbits of period up to $2^n$, but monotone systems cannot
have orbits of size larger than ${n \choose \lfloor{n/2}\rfloor}
\approx 2^n \sqrt{\frac{2}{n\pi}}$; these are all classical facts
in Boolean circuit design~\cite{Gilbert54}. It is also known that
the upper bound is tight \cite{Gilbert54}, in the sense that it is
possible to construct Boolean systems on $n$ variables, made up of
monotone functions, for which orbits of the maximal size ${n
\choose \lfloor{n/2}\rfloor}$ given by Sperner's Theorem exist.
This number is still exponential in $n$. However, anecdotal
experience suggests that monotone systems constructed according to
reasonable interconnection topologies and/or using restricted
classes of gate functions, tend to exhibit shorter orbits
\cite{Predrag,GreilDrossel}. One may ask if the
\emph{architecture} of the network, that is, the structure of its
dependency (also called interconnection) graph, helps insure
shorter orbits.  In this direction, the paper \cite{aracena-1}
showed that on certain graphs, called there ``caterpillars'',
monotone networks can only have cycles of length at most two in
their phase space.

The present paper asks the even more general question of whether networks that
are not necessarily made up from monotone functions, but which are ``close to
monotone'' (in a sense to be made precise, roughly meaning that there are few
independent
negative loops) have shorter cycles than networks which are relatively farther
to monotone.

In \cite{ejc05es}, we conjectured that ``smaller distance to
monotone'' should correlate with more ordered (less ``chaotic'') behavior,
for random Boolean networks.
A partial confirmation of this conjecture was provided in \cite{Kwon-Cho},
where the relationship between the dynamics of random Boolean networks and the
ratio of negative to positive feedback loops was investigated, albeit only for
the special case of small Kauffman-type NK and NE networks, and with the
additional restriction that all nodes
have the same function chosen from AND, OR, or UNBIAS.
Based on computer simulations, the authors of \cite{Kwon-Cho} found a
positive (negative) correlation between the ratio of fixed points (other
limit cycles) and the ratio of positive feedback loops.
Observe that this differs from our conjecture in two fundamental ways:
(1) our measure of disorder is related to the number of ``independent''
negative loops, rather than their absolute number, and (2) we do not
consider that the number of positive loops should be part of this measure: a
large number of negative loops will tend to produce large periodic orbits,
even if the negative to positive ratio is small due to a larger number of
positive loops.

\begin{figure}[bpth]
\begin{center}
\includegraphics[width=2in]{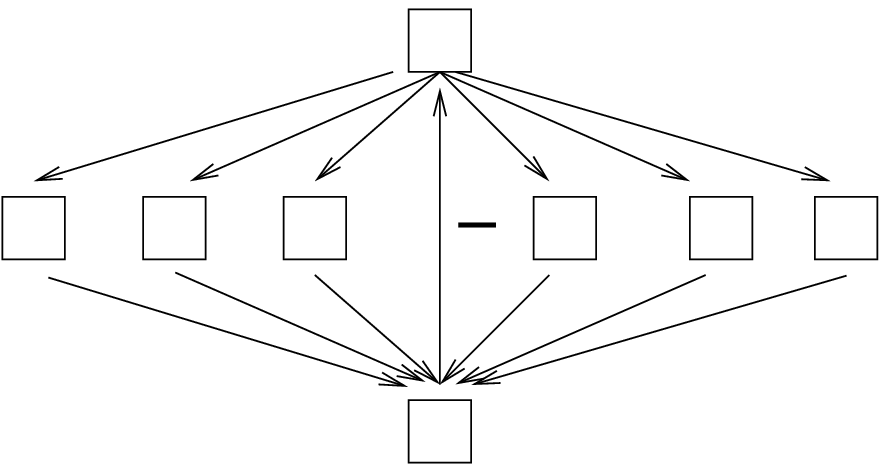}
\caption{}
\label{fig:fannet}
\end{center}
\end{figure}

Thus, in the spirit of the conjecture in \cite{ejc05es}, the current paper has
as its goal an experimental study of the effect of independent negative
feedback loops on network dynamics, based on an appropriately defined measure
of \emph{distance to positive-feedback}.
We study the effect of this distance on features of the network dynamics,
namely the number and length of limit cycles.
Rather than focusing on the number of
negative feedback loops in the network, as the characteristic
feature of a network, we focus on the number of switches of the
activation/inhibition character of edges that need to be made in
order to obtain a network that has only positive feedback loops.
We relate this measure to the cycle structure of the phase space
of the network. It is worth emphasizing that the absolute number of negative
feedback loops and the distance to positive feedback are not
correlated in any direct way, as it is easy to construct
networks with a fixed distance to positive feedback that have
arbitrarily many negative feedback loops, see Figure \ref{fig:fannet}.

\subsection*{Motivations}

There are three different motivations for posing the question that
we ask in this paper. The first is that most biological networks
appear to have highly regular dynamical behavior, settling upon
simple periodic orbits or steady states. The second motivation is
that it appears that real biological networks such as gene
regulatory networks and protein signaling networks are indeed
close to monotone
\cite{biosystems06,almostmonotone_journal,maayan07}. Thus, one may
ask if being close to monotone correlates in some way with shorter
cycles. Unfortunately, as mentioned above, one can build networks
that are monotone yet exhibit exponentially long orbits.  This
suggests that one way to formulate the problem is through a
statistical exploration of graph topologies, and that is what we
do here.

A third motivation arises from the study of systems with
continuous variables, which arguably provide more accurate models
of biochemical networks. There is a rich theory of
continuous-variable monotone (to be more precise, ``cooperative'')
systems. These are systems defined by the property that an
inequality $\textbf{a}(0)<\textbf{b}(0)$ in initial conditions
propagates in time so that the inequality
$\textbf{a}(t)<\textbf{b}(t)$ remains true for all future times
$t>0$. Note that this is entirely analogous to the Boolean case,
when one makes the obvious definition that two Boolean vectors
satisfy the inequality $\textbf{a}=(a_1,\ldots,a_n)\leq
\textbf{b}=(b_1,\ldots,b_n)$ if $a_i\leq b_i$ for each
$i=1,\ldots,n$ (setting $0<1$). Monotone continuous systems have
convergent behavior.  For example, in continuous-time (ordinary
differential models), they cannot admit any possible stable
oscillations \cite{hadeler83,hirschAMS84,Hirsch-Smith}, and, when
there is only one steady state, every bounded solution converges
to this unique steady state (monostability), see Dancer
\cite{dancer98}.  When, instead, there are multiple steady-states,
the Hirsch Generic Convergence Theorem
\cite{Hirsch,Hirsch2,smith,Hirsch-Smith} is the fundamental
result; it states, under an additional technical assumption
(``strong'' monotonicity) that generic bounded solutions must
converge to the set of steady states.  For discrete-time strongly
monotone systems, generically also stable oscillations are allowed
besides convergence to equilibria, but no more complicated
behavior. In neither case, discrete-time or continuous-time
continuous monotone systems, one observes ``chaotic'' behavior. It
is an open question whether continuous systems that are in some
sense close to being monotone have more regular behavior, in a
statistical sense, than systems that are far from being monotone,
just as for the Boolean analog considered in this paper. The
Boolean case is more amenable to computational exploration than
continuous-variable systems, however. Since long orbits in
discrete systems may be viewed as an analog of chaotic behavior,
we focus on lengths of orbits.


One can proceed in several ways to define precisely the meaning of
distance to positive feedback.  One associates to a network made
of unate (definition below) gate functions a signed graph whose
edges have signs (positive or negative) that indicate how each
variable affects each other variable (activation or inhibition).
The first definition, explored in
\cite{ejc05es,dasgupta_enciso05,biosystems06,huffnerWEA07,almostmonotone_journal}
starts from the observation that in a network with all monotone
node functions there are no negative undirected cycles.
Conversely, if the dependency graph has no undirected negative
parity cycles (a ``sign-consistent'' graph), then a change of
coordinates (globally replacing a subset of the variables by their
complements) renders the overall system monotone. Thus, asking
what is the smallest number of sign-flips needed to render a graph
sign-consistent is one way to define distance to monotone. This
approach makes contact with areas of statistical physics (the
number in question amounts to the ground energy of an associated
Ising spin-glass model), as well as with the general theory of
graph-balancing for signed graphs \cite{zaslavsky98} that
originated with Harary~\cite{harary1953}. It is also consistent
with the generally accepted meaning of ``monotone with respect to
some orthant order'' in the ODE literature as a system that is
cooperative under some inversion of variables.

A second, and different, definition, starts from the fact that a
network with all monotone node functions has, in particular, no
negative-sign \emph{directed} loops. For a strongly connected
graph, the property that no directed negative cycles exist is
equivalent to the property that no undirected negative cycles
exist. However, for non-strongly connected graphs, the properties
are not the same. Thus, this second property is weaker. The second
property is closer to what biologists and engineers mean by ``not
having negative feedbacks'' in a system, and hence is perhaps more
natural for applications. In addition, it is intuitively clear
that negative feedbacks should be correlated to possible
oscillatory behavior. (This is basically Thomas' conjecture. See
\cite{almostmonotone_journal} for precise statements for
continuous-time systems; interestingly, published proofs of
Thomas' conjecture use the first definition, because they appeal
to results from monotone dynamical systems.) Thus, one could also
define distance to monotone as the smallest number of sign-flips
needed to render a graph free of negative directed loops. To avoid
confusion, we will call this notion, which is the one studied in
this paper, {\it distance to positive-feedback}, or just
``PF-distance''.


\section*{Theory}

\subsection*{Distance to positive-feedback}

We give here the basic definitions of the concepts relevant to the
study.

\begin{definition}
\label{mono-unate}
Let $k = \{0, 1\}$ be the field with two elements. We order the two
elements as $0 < 1$.  This ordering can be extended to a partial
ordering on $k^n$ by comparing vectors coordinate-wise in the
lexicographic ordering.
\begin{enumerate}
\item A Boolean function $h: k^n \longrightarrow k$ is {\it
monotone} if, whenever $\mathbf a \leq \mathbf b$ coordinate-wise,
for $\mathbf a, \mathbf b \in k^n$, then $h(\mathbf a)\leq
h(\mathbf b)$.
\item A Boolean function $h$ is {\it unate} if, whenever $x_i$ appears in $h$,
the following holds:  Either
\begin{enumerate}
\item For all $a_1,\ldots,a_{i-1},a_{i+1},\ldots , a_n\in k$, \\
$h(a_1,\ldots ,a_{i-1},0,a_{i+1},\ldots ,a_n) \leq$  \\
\vspace{.3cm} \hspace{2cm} $h(a_1,\ldots,a_{i-1},1,a_{i+1},\ldots ,a_n)$, or
\item For all $a_1,\ldots,a_{i-1},a_{i+1},\ldots , a_n\in k$, \\
$h(a_1,\ldots ,a_{i-1},0,a_{i+1},\ldots ,a_n) \geq $ \newline
\vspace{.3cm} \hspace{2cm} $h(a_1,\ldots,a_{i-1},1,a_{i+1},\ldots
,a_n)$.
\end{enumerate}
\end{enumerate}
The definition of unate function is equivalent to requiring that
whenever $ a_i$ appears in $h$, then it appears either everywhere
as $a_i$ or everywhere as $\neg a_i:= 1+a_i$.
\end{definition}

Let $f$ be a Boolean network with variables $x_1, \ldots ,x_n$,
 and coordinate functions $f_1, \ldots ,f_n$. That is, $f =
(f_1,\dots,f_n): k^n \longrightarrow k^n$. We can associate to $f$
its \emph{dependency graph} $\mathcal D(f)$: The vertices are
$v_1, \ldots , v_n$, corresponding to the variables
$x_1,\dots,x_n$, and there is an edge $v_i \rightarrow v_j$ if and
only if $x_i$ appears in $f_j$. If all coordinate functions $f_i$
of $f$ are unate, then the dependency graph of $f$ is a signed
graph. Namely, we associate to an edge $v_i\rightarrow v_j$ a
``+'' if $f_j$ preserves the ordering as in 2(a) of
Definition \ref{mono-unate} and a ``-'' if it reverses
the ordering as in 2(b) of  Definition \ref{mono-unate}.
For later use we observe that this graph (as any
directed graph) can be decomposed into a collection of strongly
connected components, with edges between strongly connected
components going one way but not the other. (Recall that a
strongly connected directed graph is one in which any two vertices
are connected by a directed path.)  That is, the graph can be
represented by a partially ordered set in which the strongly
connected components make up the elements and the edge direction
between components determines the order in the partially ordered
set.

\begin{definition}
Let $f$ be a Boolean network with unate Boolean functions and
$\mathcal D(f)$ be its signed dependency graph. Then
\begin{enumerate}
\item $f$ is a  \emph{positive-feedback network (PF)} if $\mathcal D(f)$ does not contain
any odd parity directed cycles. (The parity of a directed cycle is
the product of the signs of all the edges in the cycle.)
\item The \emph{PF-distance} of $f$ is the smallest number of signs that
need to be changed in the dependency graph to obtain a PF network.
We denote this number by $|\mathcal D(f)|$ or simply $|f|$.
\end{enumerate}
\end{definition}
Notice that for a given
directed graph $G$, different assignments of sign to the edges
produce graphs with varying PF-distance.  In particular, there is
a maximal PF-distance that a given graph topology can support.

The dynamics of $f$ are presented in a directed graph, called the
\emph{phase space} of $f$, which has the $2^n$ elements of $k^n$
as a vertex set, and there is an edge $\mathbf a\rightarrow
\mathbf b$ if $f(\mathbf a) = \mathbf b$.  It is straightforward
to see that each component of the phase space has the structure of
a directed cycle, a {\it limit cycle}, with a directed tree
feeding into each node of the limit cycle.  The elements of these
trees are called {\it transient states}.

In this paper we relate the dynamics of a Boolean network to its PF-distance. The following
is a motivational example that explains the main results.

\begin{figure*}[tbhp]
\centerline{ \raise25pt\hbox{ \framebox{
\includegraphics[width=0.12\textwidth]{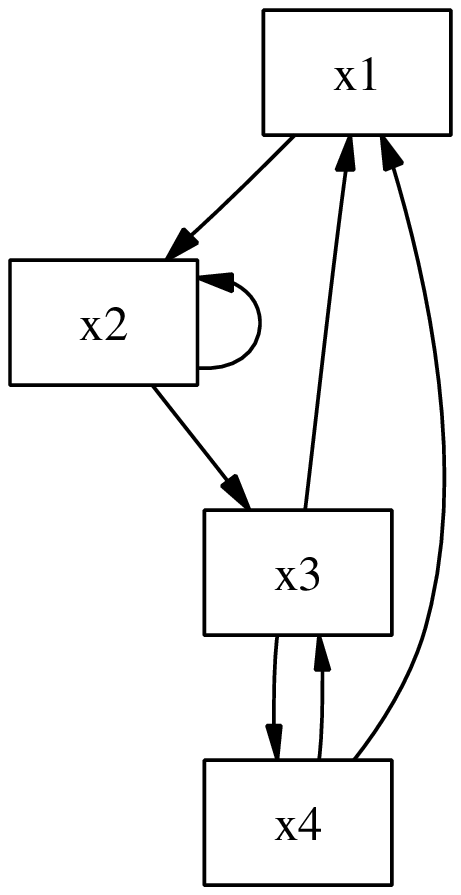}}}
\quad \quad  \raise30pt\hbox{ \framebox{
\includegraphics[width=0.45\textwidth]{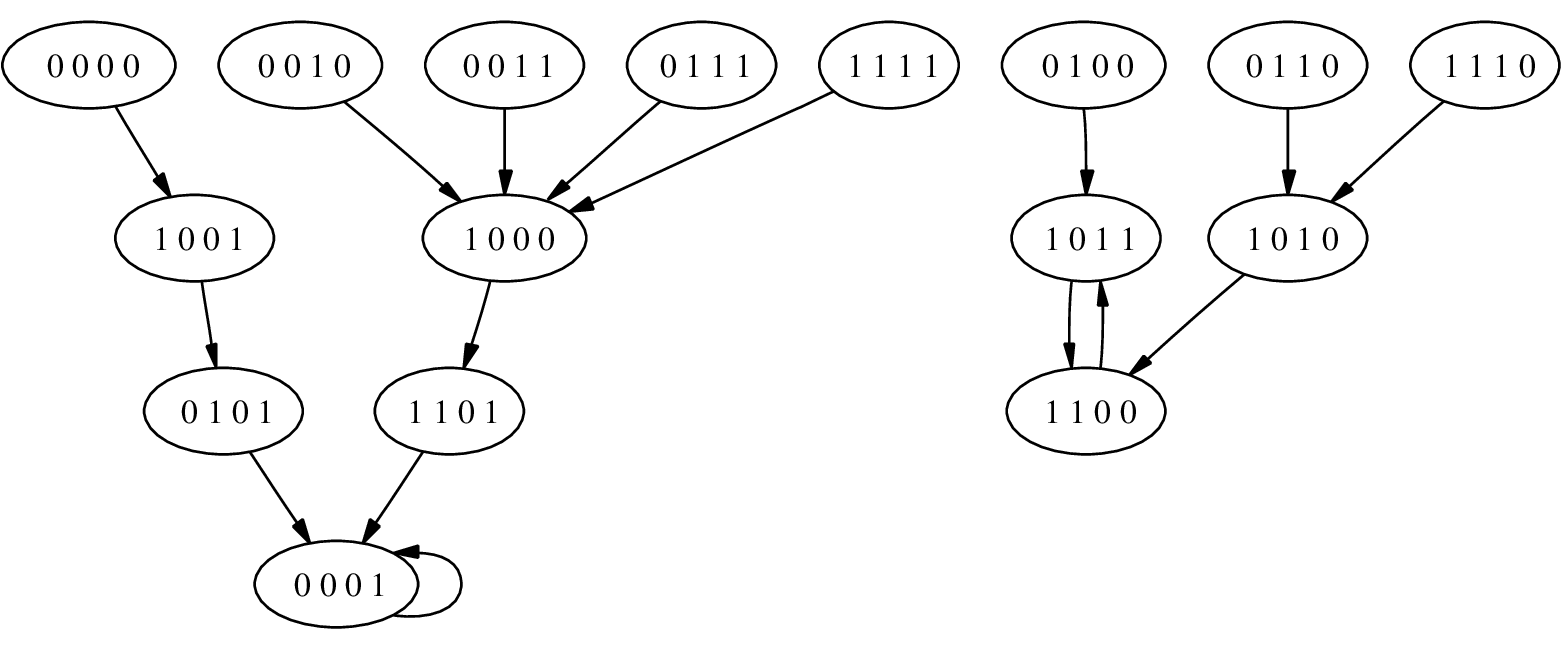}}}
\quad \quad  \raise10pt\hbox{ \framebox{
\includegraphics[width=0.30\textwidth]{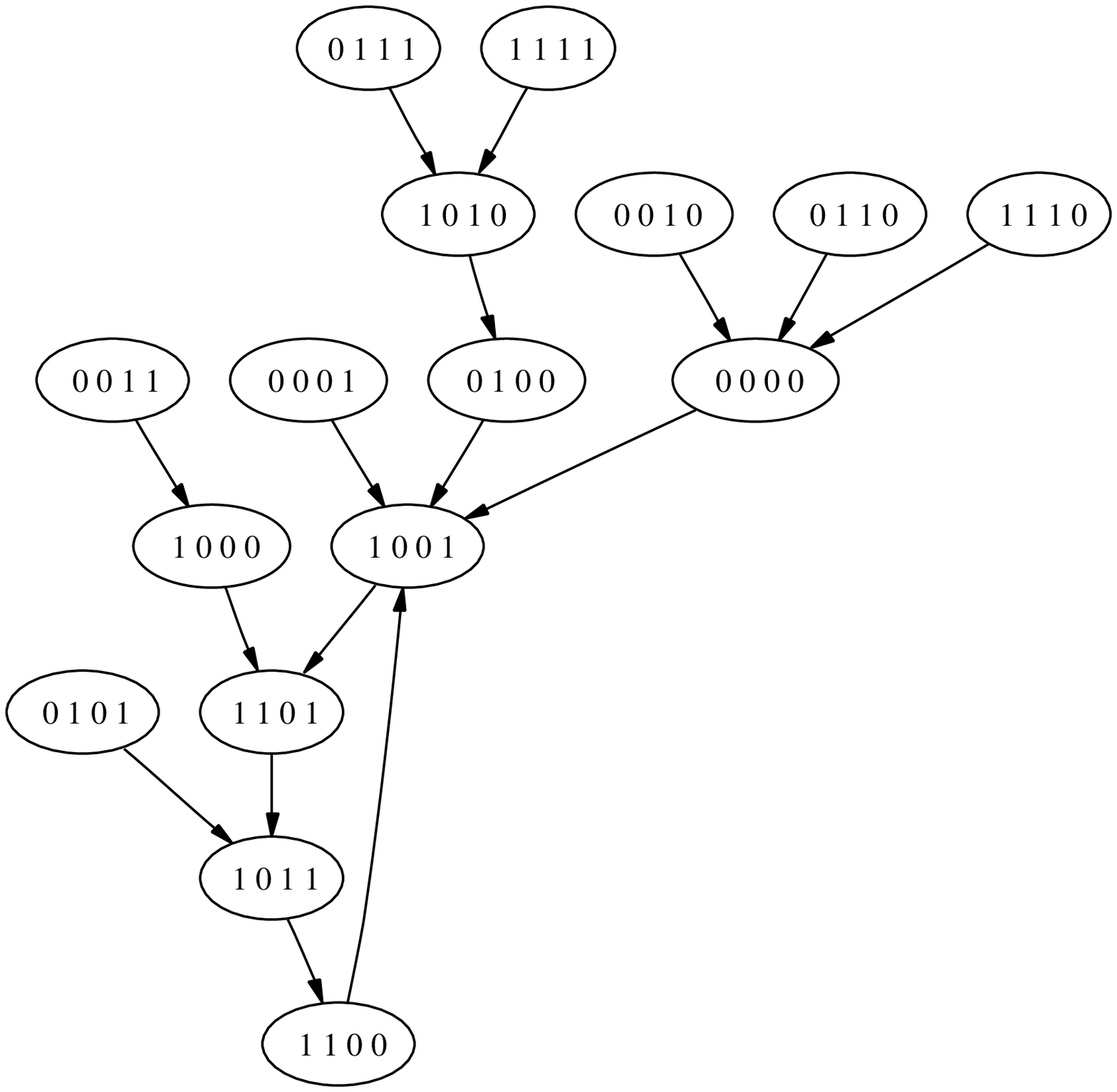}}}}
\caption{}
\label{fig-ex}
\end{figure*}

\begin{figure}[bpth]
\begin{center}
\includegraphics[width=4in]{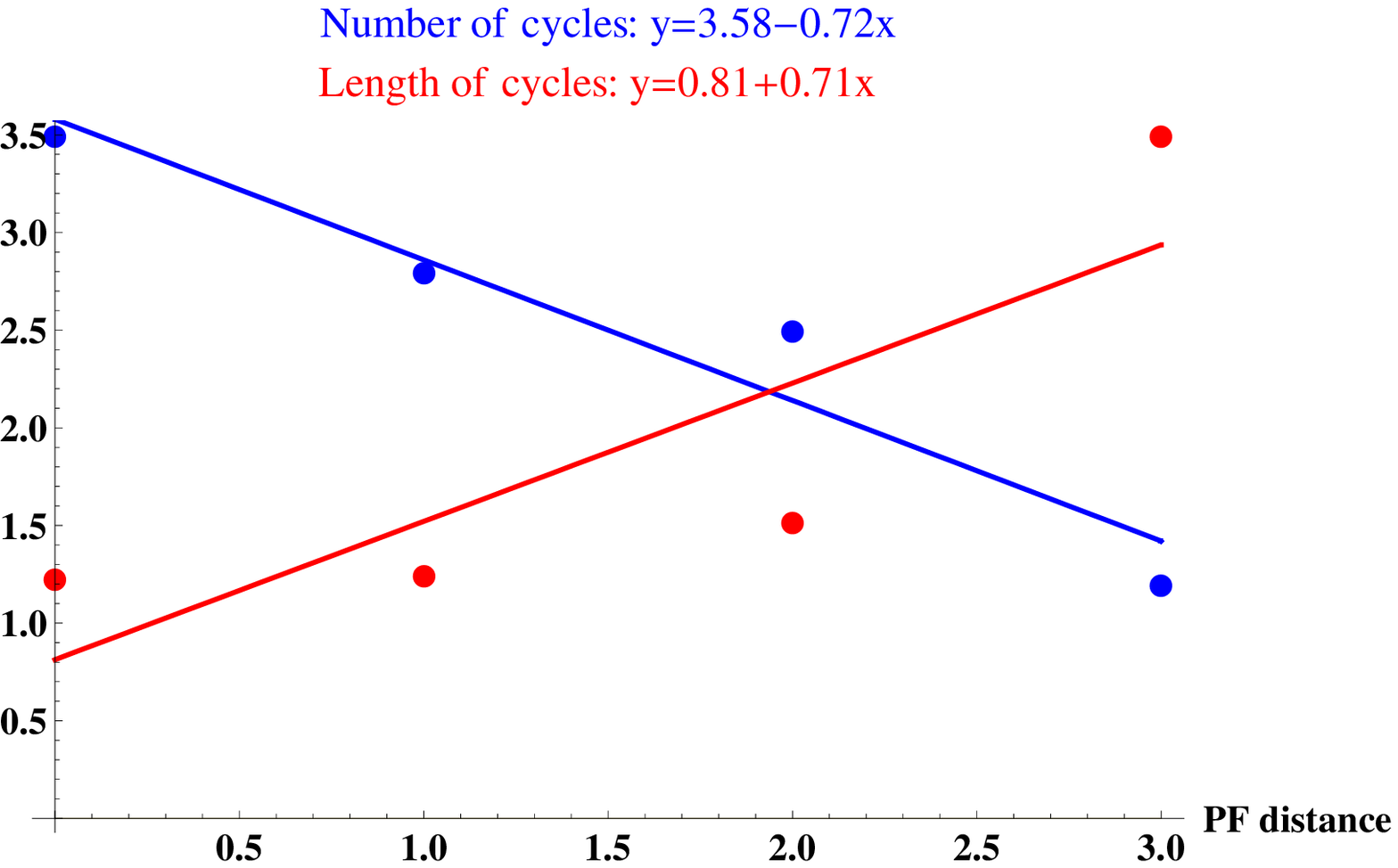}
\caption{}
\label{slopes}
\end{center}
\end{figure}

\begin{example}
\label{run_exam} Let $G$ be the directed graph depicted in Fig.
\ref{fig-ex} (left). It is easy to check that the maximal
PF-distance of $G$ is 3. Let $f=(x_3 \vee \neg x_4, x_1 \wedge
x_2, x_2 \wedge \neg x_4, \neg x_3): \{0,1\}^4 \longrightarrow
\{0,1\}^4$ and $g = (\neg x_3 \vee x_4, x_1 \wedge \neg x_2, x_2
\wedge x_4, \neg x_3): \{0,1\}^4 \longrightarrow \{0,1\}^4$. It is
clear that $f$ and $g$ are sign-modifications of the same PF
network $(x_3 \vee x_4, x_1 \wedge x_2, x_2 \wedge x_4, x_3)$, in
particular, they have the same (unsigned) dependency graph.
However, the PF-distance of $f$ is 0 while it is 3 for $g$. The
phase space of $f$  is depicted in Fig. \ref{fig-ex} (middle) and
that of $g$ is on the right. Notice that $f$ has two limit cycles
of lengths $1$ and $2$, respectively, while $g$ has only one limit
cycle of length $4$.

For each distance $0\leq d \leq 3$, we analyze the dynamics of  10
random PF networks and their sign modifications of distance $d$ on
the directed graph in Fig. \ref{fig-ex} (left). The average of the
numbers (lengths) of limit cycles is computed as in Table
\ref{av_tab}. The best fit-line of the averages of the number
(length) of limit cycles is computed and its slope is reported as
in Fig. \ref{slopes}. The details of this analysis are provided in
the \emph{Supporting Inforamtion}.
\end{example}

We have repeated the experiment in Example \ref{run_exam} above
for many different graphs and observed that the slope of the best
fit-line of the length (resp. number) of limit cycles is  positive
(resp. negative) most of the time. In the methods section we
present the details of the experiment and the algorithms used in
the computations. The results of these experiments are described
next.

\section*{Methods}




The main results of this paper relate the PF-distance of Boolean
networks with the number and length of their limit cycles.
Specifically, our hypothesis is that, for Boolean networks
consisting of unate functions, \emph{as the PF-distance
increases, the total number of limit cycles decreases on average
and their average length increases}.
This is equivalent to saying that for most or all
experiments this slope is negative for the number of limit cycles
and is positive for their length.

To test this hypothesis we analyzed the dynamics of
more than six million Boolean networks arranged in
about 130,000 experiments
on random graphs with the number of nodes 5, 7, 10, 15, 20, or 100
and maximum in-degree 5 for each node.

\subsection*{Random generation of unate functions}
We generated a total of more than 130,000 random directed graphs,
where each graph has 5,7,10,15, 20, or 100 nodes,
with maximum in-degree 5 for each node. The
graphs were generated as random adjacency matrices, with the
restriction that each row has at least one 1 and at most five 1's.
For each graph directed $G$, we generated 10 Boolean networks with
unate functions and dependency graph $G$, by using the following fact.

\begin{lemma}
A Boolean function $f$ of $n$ variables is unate if and only if it
is of the form $f(\mathbf x) = g(\mathbf x+\mathbf s)$, where $g$
is a monotone Boolean function of $n$ variables and $\mathbf s\in
k^n$ and ``+'' denotes addition modulo 2.
\end{lemma}
\noindent
\textbf{Proof}.
If $f$ is unate then each variable $x_i$ appears in $f$ always as
$x_i$ or always as $\neg x_i$.  Suppose that all $x_i$ appear
without negations.  Then $f$ is constructed using $\wedge$ and
$\vee$.  Hence $f$ is monotone.  Otherwise, let $\mathbf s\in k^n$
be the vector whose $i$th entry is 1 if and only if $x_i$ appears
as $\neg x_i$ in $f$.  Then $g(\mathbf x) = f(\mathbf x+\mathbf
s)$ is a monotone function and $f(\mathbf x) = g(\mathbf x+\mathbf
s)$.  The converse is clear.

\smallskip

So in order to generate unate functions it is sufficient to generate
monotone functions.  We generated the set $M_i$ of monotone
functions in $i$ variables by exhaustive search for $i=1,\ldots, 5$.
(For example, $M_5$ has 6894 elements.)  Unate functions for a
given signed dependency graph can then be generated by choosing
random functions from $M_i$ and random vectors $\mathbf s\in k^n$.
The nonzero entries in $\mathbf s$ for a given node correspond to
the incoming edges with negative sign in the dependency graph. Using
this process we generated Boolean networks with unate Boolean
functions.

We then carry out the following steps.

\subsection*{The Experiment}
Let $G$ be a random unsigned directed graph on $n$ nodes with a maximal PF-distance $t$, and let $D \leq t$.
Consider $10$ unate Boolean networks chosen at random with $G$ as their dependency graph.
\begin{enumerate}
\item For $1 \leq d \leq D$, let $G_d$ be a signed graph of $G$ of distance $d$.
    \begin{enumerate}
    \item For each network $f$ of the ten networks,
        \begin{enumerate}
        \item Let $g$ be a modified network of $f$ such that  $\mathcal D(g) = G_d$; the signed dependency graph of $g$ is $G_d$.
        \item Compute the number and length of all limit cycles in the phase space of $g$.
        \end{enumerate}
    \item Compute the average number $N$ (resp. average length $L$) of limit cycles in the phase spaces of the $g's$.
    \end{enumerate}
\item Compute the slope $s_N$ (resp. $s_L$) of the best fit-line
of the $N's$ (resp. $L's$).
\end{enumerate}

\subsection*{Computation of PF-distance}
Let $f$ be a Boolean network with unate Boolean
functions and let $|f|$ be its PF-distance.
The proofs of the following facts are straightforward.
\begin{enumerate}
\item Suppose the dependency graph of $f$ has a negative feedback
loop at a vertex. Let $f'$ be the Boolean network obtained by
changing a single sign to make the loop positive.  Then $|f|=
|f'|+1$. \item Let $H_1,\ldots H_s$ be the strongly connected
components of the dependency graph $D(f)$. Then
$$
|\mathcal D(f)| = \sum_{i=1}^s |H_i|.
$$
\end{enumerate}
The algorithm for computing $|\mathcal D(f)|$ now follows.

\framebox{
\begin{minipage}[l]{3.0in}
\hspace{2pt}
\textbf{Algorithm: Distance to PF}

\begin{tabbing}
\hspace{10pt}\=\textbf{Input}: \hspace{9pt}\= A signed, directed graph $G$. \\
\> \textbf{Output}: \>  $|G|$; the PF-distance of $G$.
\end{tabbing}
Let $d = 0$.
    \begin{enumerate}
        \item Let $G_1, \ldots, G_r$ be the collection of all signed
        graphs obtained by making exactly $d$ sign changes in $G$. \label{step1}
        \item For $i=1,\dots, r$ \\
            If $G_i$ is PF, then RETURN $|G| = d$. \label{is_PF}
        \item Otherwise, $d := d+1$, Go to Step (\ref{step1}) above.
    \end{enumerate}
\end{minipage}
}


In Step (\ref{is_PF}) above, to check whether a strongly connected
graph is PF, it is equivalent to check whether it has any
(undirected) negative cycles, which can easily be done in many
different ways, see, e.g., \cite{almostmonotone_journal}. This
algorithm must terminate, since $G$ has finitely many edges and
hence the PF-distance of $G$ is finite.

If $G$ has $m$ directed edges, then there are $2^m$ possible sign assignments. However,
to compute the maximal PF-distance, one does not need to find the PF-distance of such possible assignments,
see \emph{Supporting Information} for the algorithm we used to compute the maximal PF-distance.

\section*{Results}

In Table \ref{total} we present the percentage of experiments
that do conform to our hypothesis for the average of number
of limit cycles as well as  for the average length
of limit cycles.
It should be mentioned here that a computationally expensive part of an
experiment is the computation of the maximal PF-distance of a given
directed graph and becomes prohibitive for even modest-size graphs,
with, e.g., 10 nodes.  So unlike in Example \ref{run_exam},
for networks on more than 5 nodes,
we only considered PF-distances that are less than or equal to the number
of nodes in the network. (See the
Methods Section for a detailed description of the experiment.)
In fact for graphs with 20 (resp. 100) nodes, all considered networks
have PF-distance less than or equal to 5 (resp. 10). We argue below that
this is the reason for the drop in  the percentage
of experiment that conform to our hypothesis as
the number of nodes increases.

\begin{figure*}[htbp]
\centerline{
\raise10pt\hbox{ \framebox{
\includegraphics[width=1.5in]{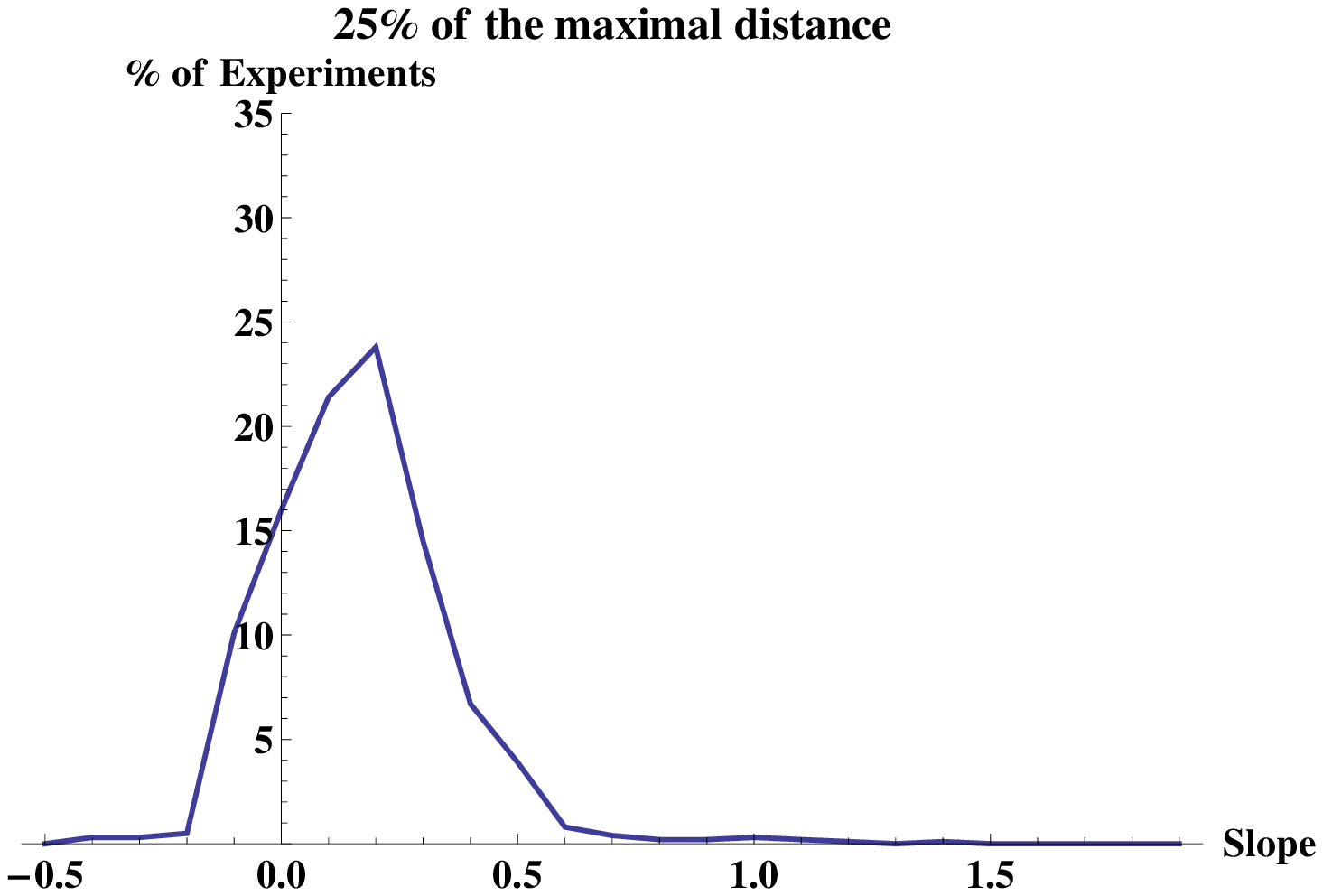}}}
\raise10pt\hbox{ \framebox{
\includegraphics[width=1.5in]{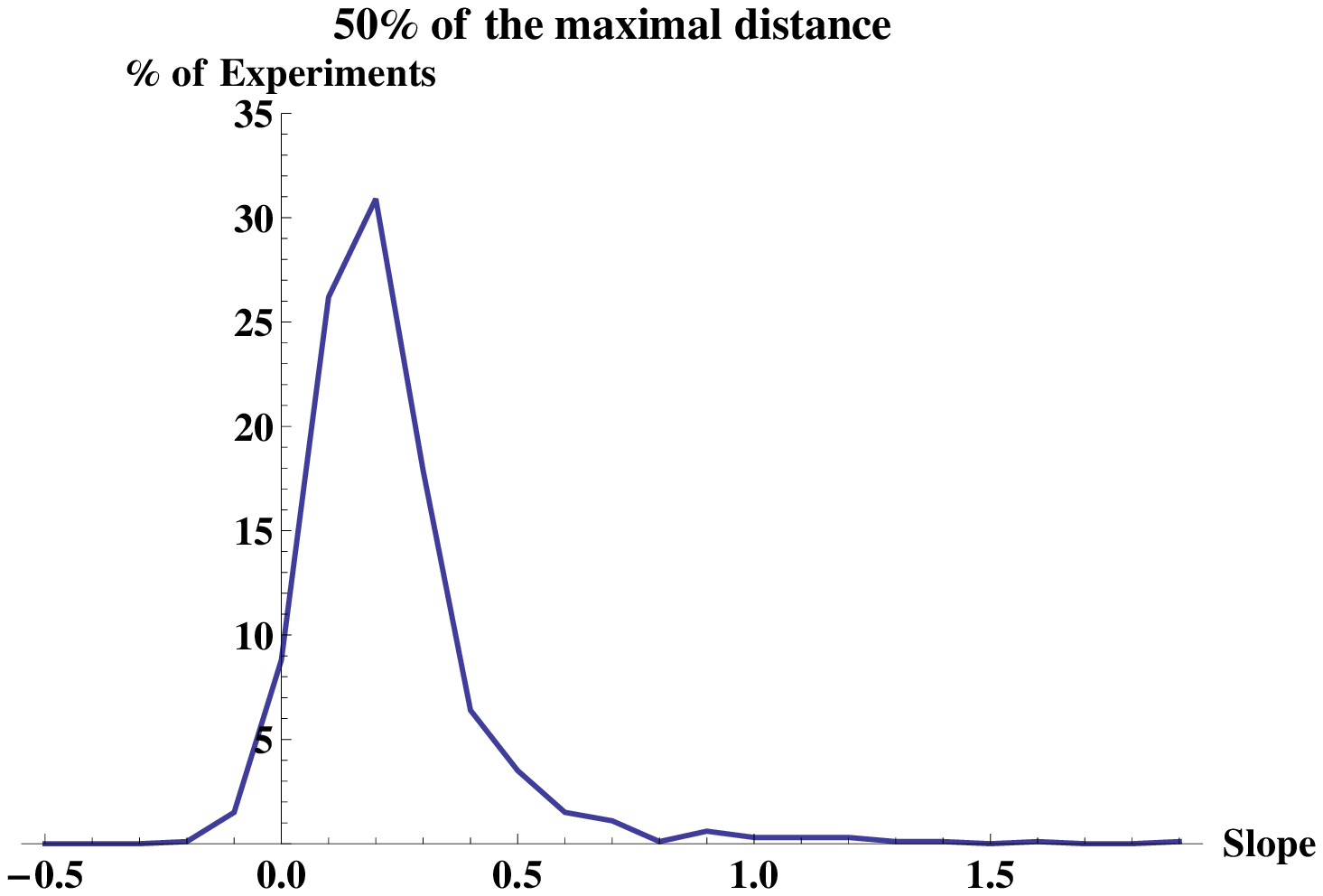}}}
\raise10pt\hbox{ \framebox{
\includegraphics[width=1.5in]{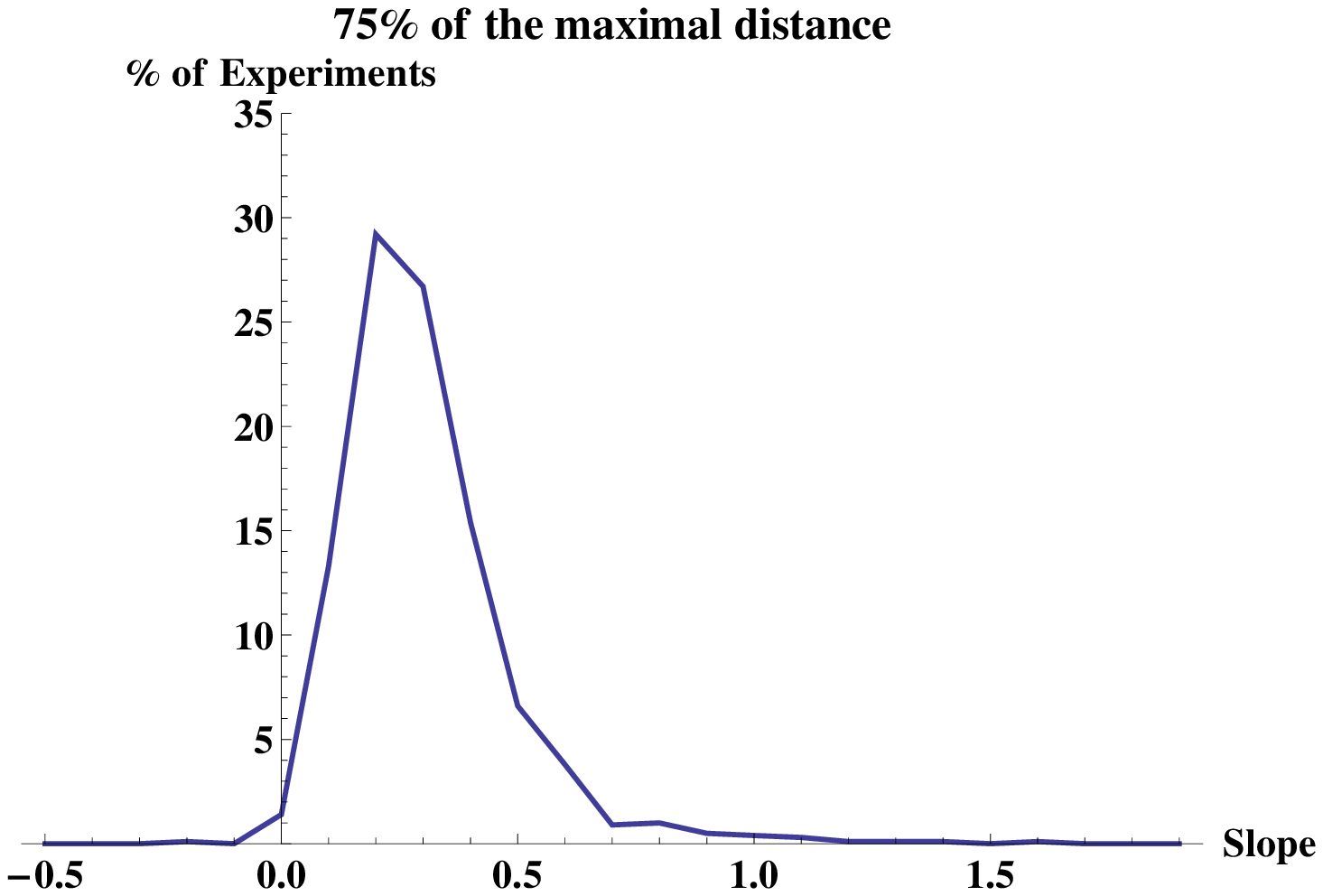}}}
\raise10pt\hbox{ \framebox{
\includegraphics[width=1.5in]{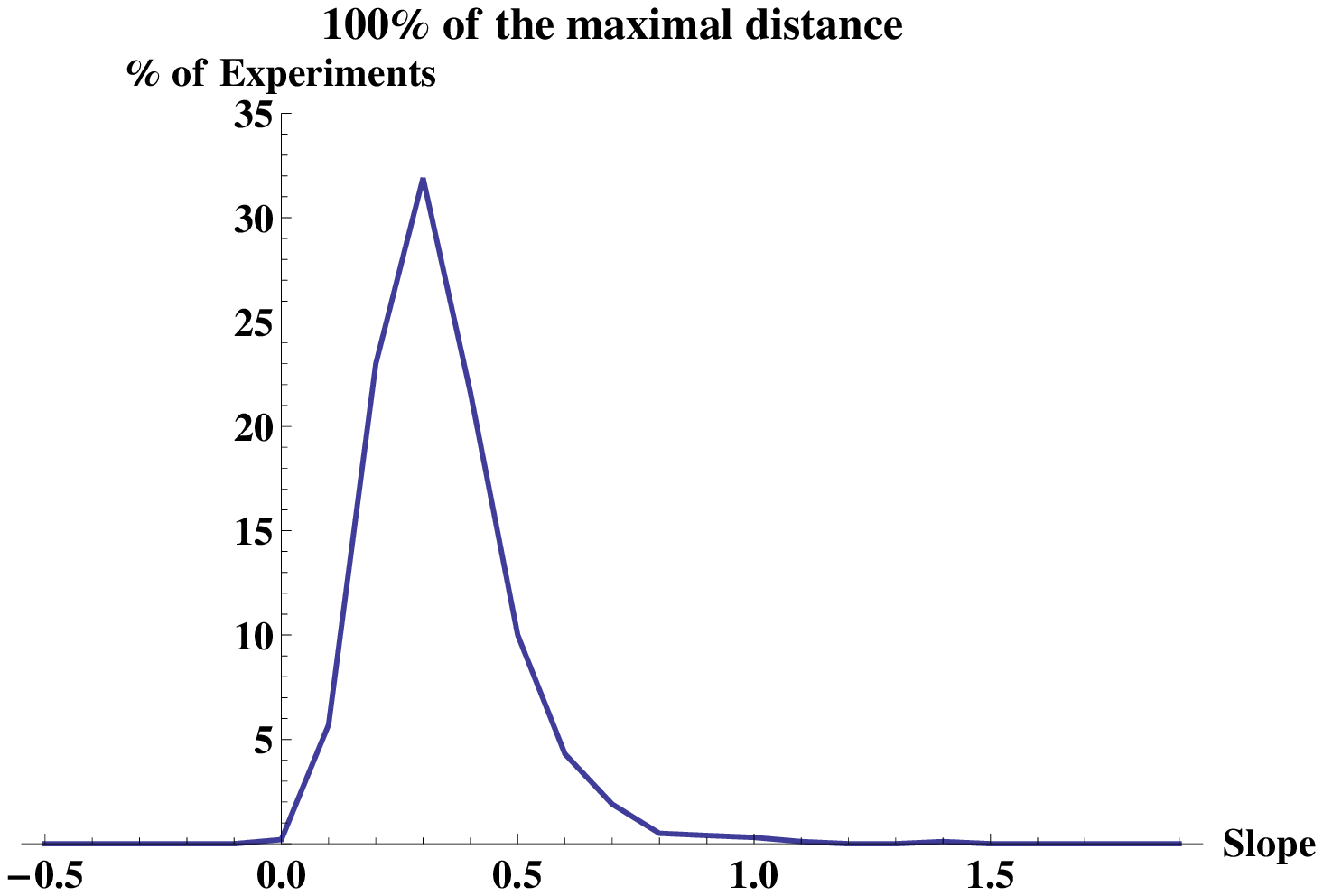}}}
}
\caption{}
\label{5-node_AL}
\end{figure*}

\begin{figure*}[htbp]
\centerline{
\raise10pt\hbox{ \framebox{
\includegraphics[width=1.5in]{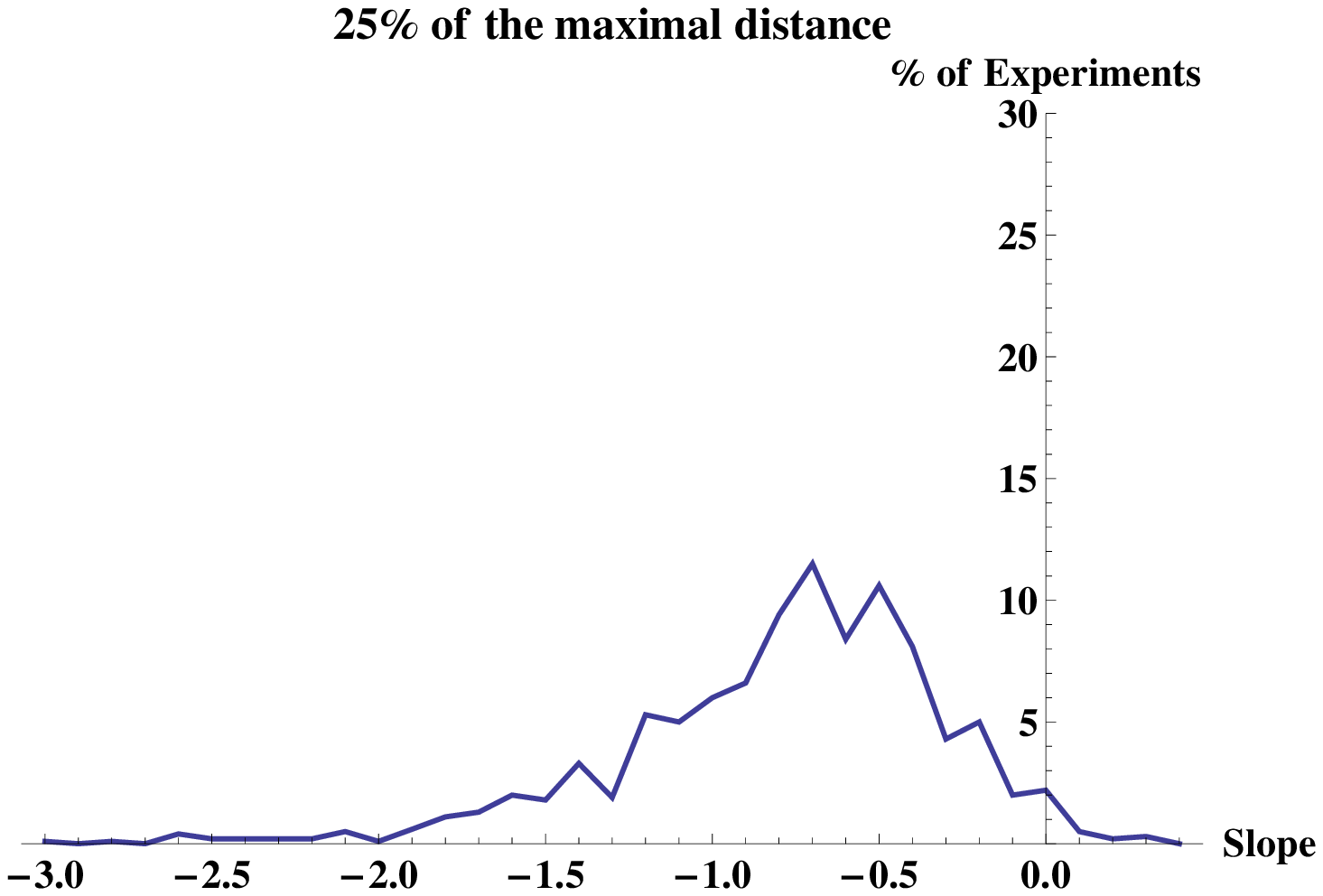}}}
\raise10pt\hbox{ \framebox{
\includegraphics[width=1.5in]{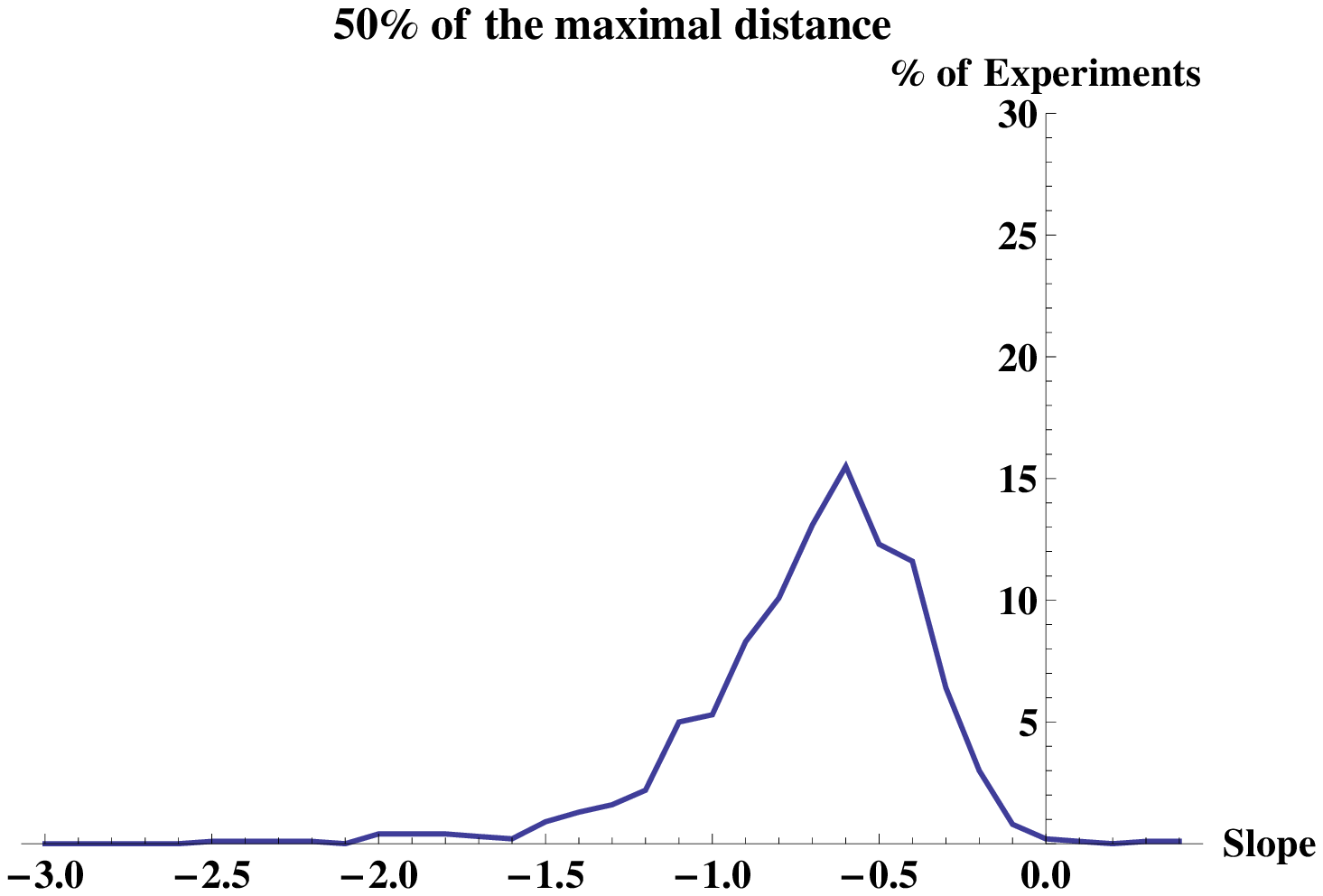}}}
 \raise10pt\hbox{ \framebox{
\includegraphics[width=1.5in]{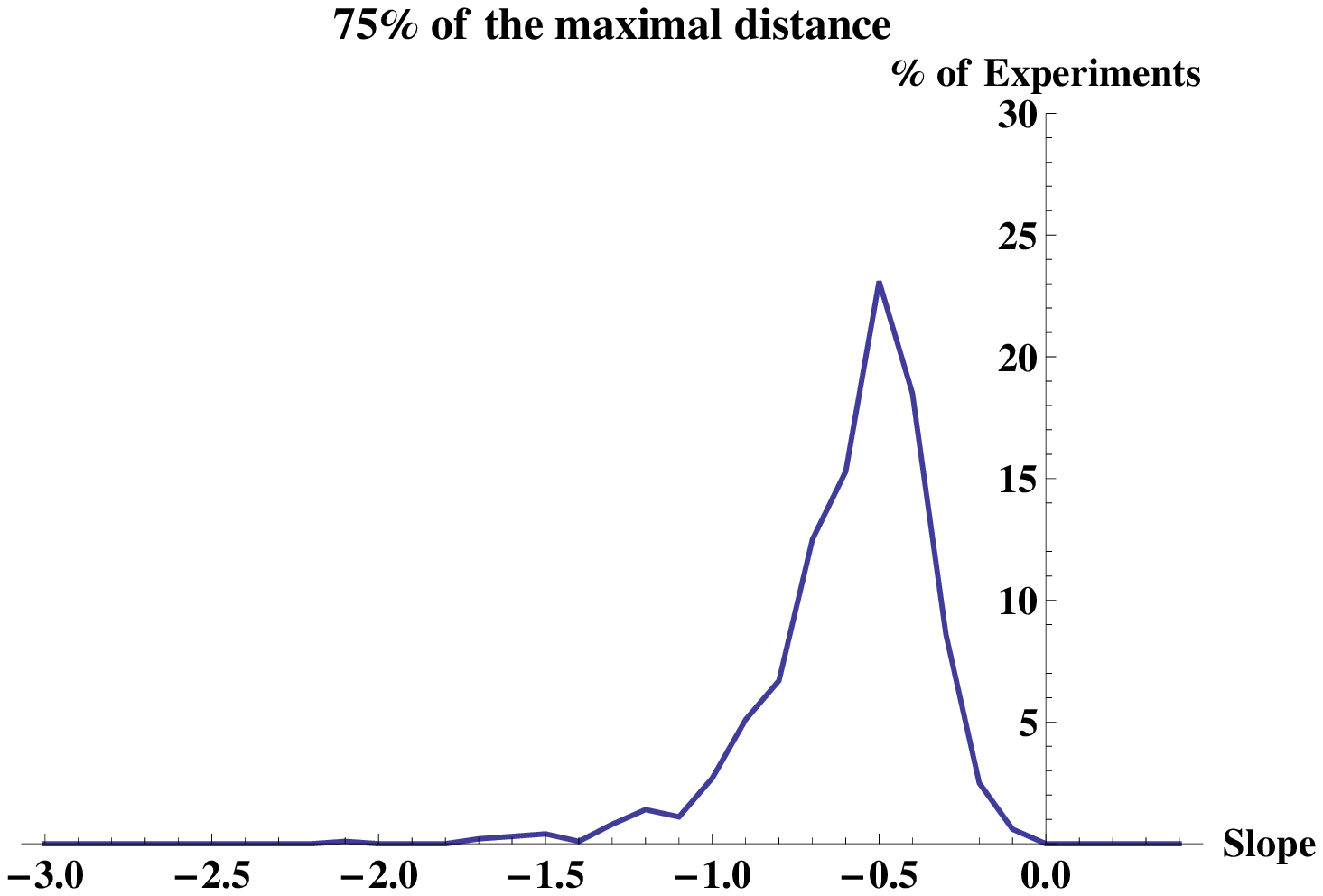}}}
\raise10pt\hbox{ \framebox{
\includegraphics[width=1.5in]{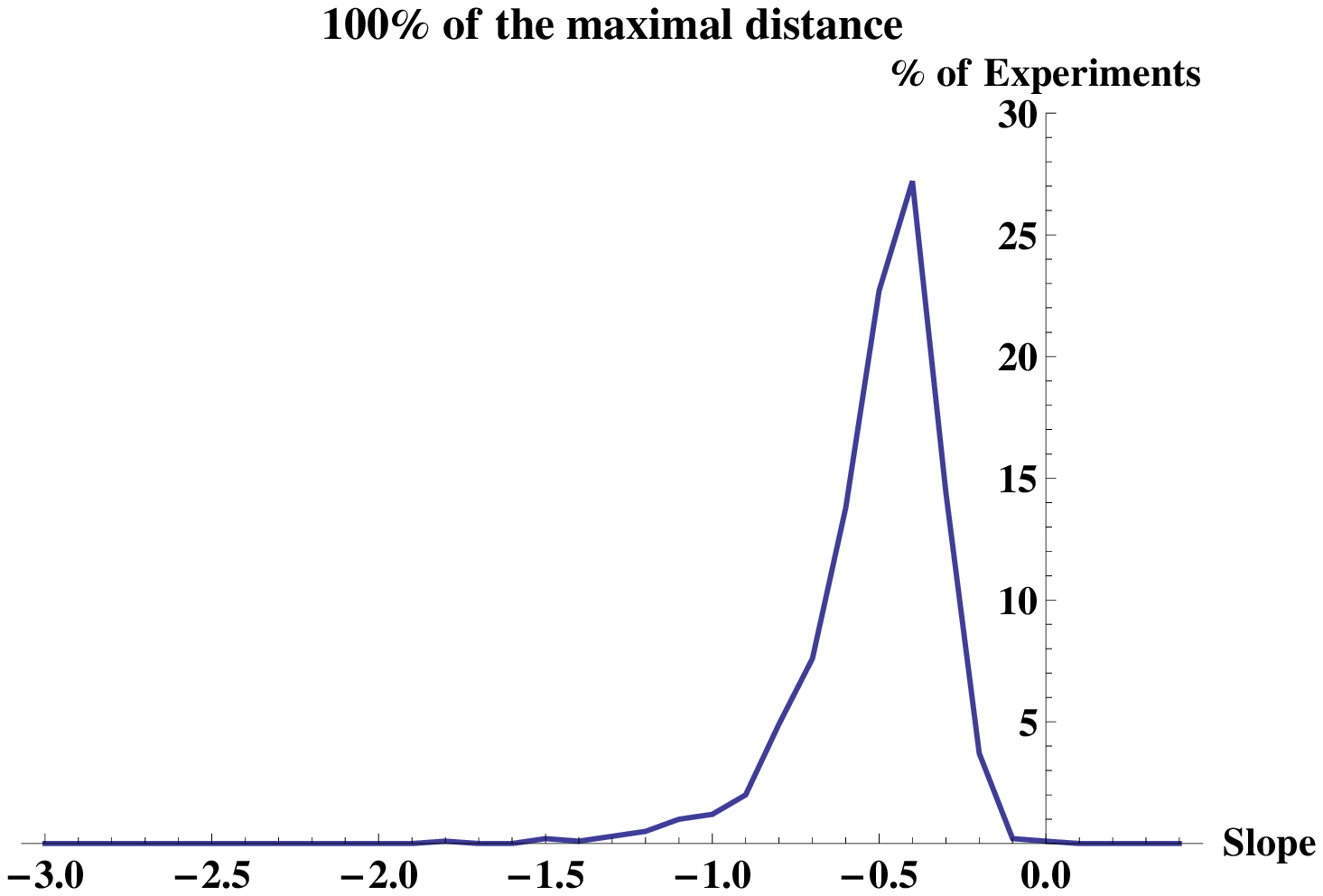}}}
}
\caption{}
\label{5-node_AN}
\end{figure*}

For networks on 5 nodes, we analyzed the dynamics of 4000
experiments by varying the PF-distance considered in the computations.
Table \ref{tab_5} shows the number of experiments that do not conform
to our hypothesis as we vary the considered PF-distance.
We also present the results in the
form of histograms, where the horizontal axis represents the slope
of the lines of best fit and the vertical axis represents the
percentage of experiments that confirm our hypothesis.
Figure \ref{5-node_AL} shows the results of the 4000 experiments
on 5-node networks.
The histograms, from left to right, show the results when PF-distance of
the network is 25\%, resp. 50\%, resp. 75\%, resp. 100\% of the
maximal PF-distance.  It can be seen in the
right-most figure that almost all experiments show positive slope of the
best-fit line, thereby conforming to the conjecture.  Similar
results for the average number of limit cycles are shown in Figure
\ref{5-node_AN}, demonstrating that if the PF-distance of networks
is allowed the whole possible range, almost all the experiments
conform to our hypothesis as we already noticed in Table \ref{tab_5}.
However, these histograms show the distribution of slopes over different
distances.

We have carried out similar computations for networks with 7 (5000
experiments)and 10 (6000 experiments) nodes; see the Supporting
information for details. The results there are not quite as clear
as for 5 node networks.  For instance, for networks with 10 nodes
(and up to 4 incoming edges per node) 31 out of 1000 experiments
did not conform to our hypothesis for networks with PF-distance up
to 5.

In summary, the exhaustive computations confirm our hypothesis
that, as the PF-distance increases, the total number of limit
cycles decreases on average and their average length increases.
Furthermore, the slopes of the best-fit lines increasingly conform
to our hypothesis the closer the PF-distance of the networks comes
to the maximum PF-distance of the network topology.


\subsection*{Supporting Information} For details of our analysis
of the 5-node networks and all other considered networks, see
\emph{Supporting Text}.

\section*{Discussion}

Negative feedback loops in biological networks play a crucial role
in controlling network dynamics.  The new measure of ``distance to
positive feedback (PF-distance)" introduced in this paper is
designed to capture the notion of ``independent" feedback loops.
We have shown that PF-distance correlates very well with the
average number and length of limit cycles in networks, key
measures of network dynamics.  By analyzing the dynamics of more than
six millions Boolean networks, we have provided evidence that
networks with a larger number of independent negative feedback
loops tend to have longer limit cycles and thus may exhibit more
``random'' or ``chaotic'' behavior. Furthermore, the number of
limit cycles tends to decrease as the number of independent
negative feedback loops increases.

In general, the problem of computing the PF-distance of a network
is NP-complete, as MAX-CUT can be mapped into it as a special
case; see
\cite{dasgupta_enciso05,biosystems06,almostmonotone_journal} for a
discussion for the analogous problem of distance to monotone. The
question of computing \emph{distance to monotone} has been the
subject of a few recent papers
\cite{dasgupta_enciso05,biosystems06,huffnerWEA07}. The first two
of these proposed a randomized algorithm based on a semi-definite
programming relaxation, while the last one suggested an efficient
deterministic algorithm for graphs with small distance to
monotone. Since a strongly connected component of a graph is
monotone if and only if it has the PF property, methods for
computing PF distance for large graphs may be developed by similar
techniques.  Work along these lines is in progress.

\smallskip
\noindent
\textbf{Acknowledgements.}
 Sontag was supported in part by NSF Grant DMS-0614371. Veliz-Cuba,
 Laubenbacher and Jarrah were  supported partially by NSF Grant
 DMS-0511441.  The computational results presented here were in part obtained
 using Virginia Tech's Advanced Research Computing
 (http://www.arc.vt.edu) 128p SGI Altix 3700 (Inferno2) system.


\bibliography{new_PF}


\clearpage
\section*{Figure Legends}

\subsubsection*{Figure~\ref{fig:fannet}.}

{A graph that has an arbitrary number of negative loops, as many as
the number of nodes in the second layer, but PF distance one: to avoid
negative feedback, it suffices to switch the sign of the single (negative)
arrow from the bottom to the top node.  All unlabeled arrows are positive.}

\subsubsection*{Figure~\ref{fig-ex}.}

{The dependency graph (left), the phase space of $f$ (middle)
  and the phase space of $g$ (right) from Example \ref{run_exam}. These graph were generated using DVD \cite{dvd}.}

\subsubsection*{Table~\ref{av_tab}.}

{The average of the numbers (lengths) of limit cycles of the networks from Example \ref{run_exam}.}

\subsubsection*{Table~\ref{total}.}

{The percentage of experiments that conform the hypotheses}

\subsubsection*{Table~\ref{tab_5}.}

{The number of experiments that did not conform to our hypothesis for  5-node networks. We considered PF-distance $25\%$, $50\%$,$75\%$, and $100\%$ of the maximal distance  . For each $d$, we considered 1000 experiments.}

\subsubsection*{Figure~\ref{5-node_AL}.}

{5-node networks. Histogram of slopes of best-fit lines to the average length of limit cycles (horizontal axis) vs. percentage of experiments with a given slope (vertical axis).  The panels from left to right include networks with increasing PF-distance, with 25\%, 50\%, 75\%, and 100\% of the maximal distance.}

\subsubsection*{Figure~\ref{5-node_AN}.}

{5-node networks. Histogram of slopes of best-fit lines to the average  number of limit cycles (horizontal axis) vs. percentage of experiments with  a given slope (vertical axis). The panels from left to right include networks with increasing PF-distance, with 25\%, 50\%, 75\%, and 100\% of the maximal distance.}

\begin{figure*}
\centerline{
\raise10pt\hbox{ \framebox{
\includegraphics[width=1.5in]{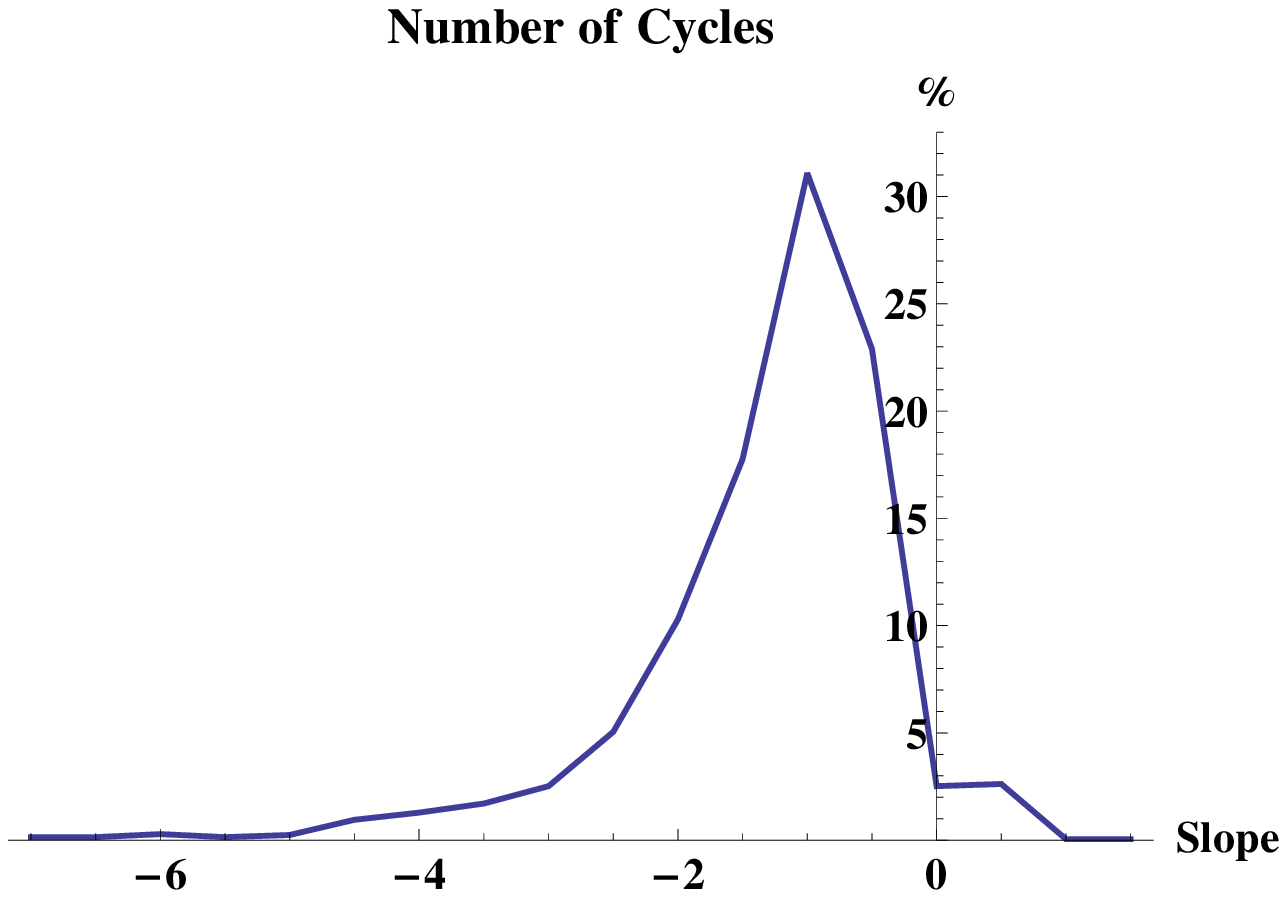}}}
 \raise10pt\hbox{ \framebox{
\includegraphics[width=1.5in]{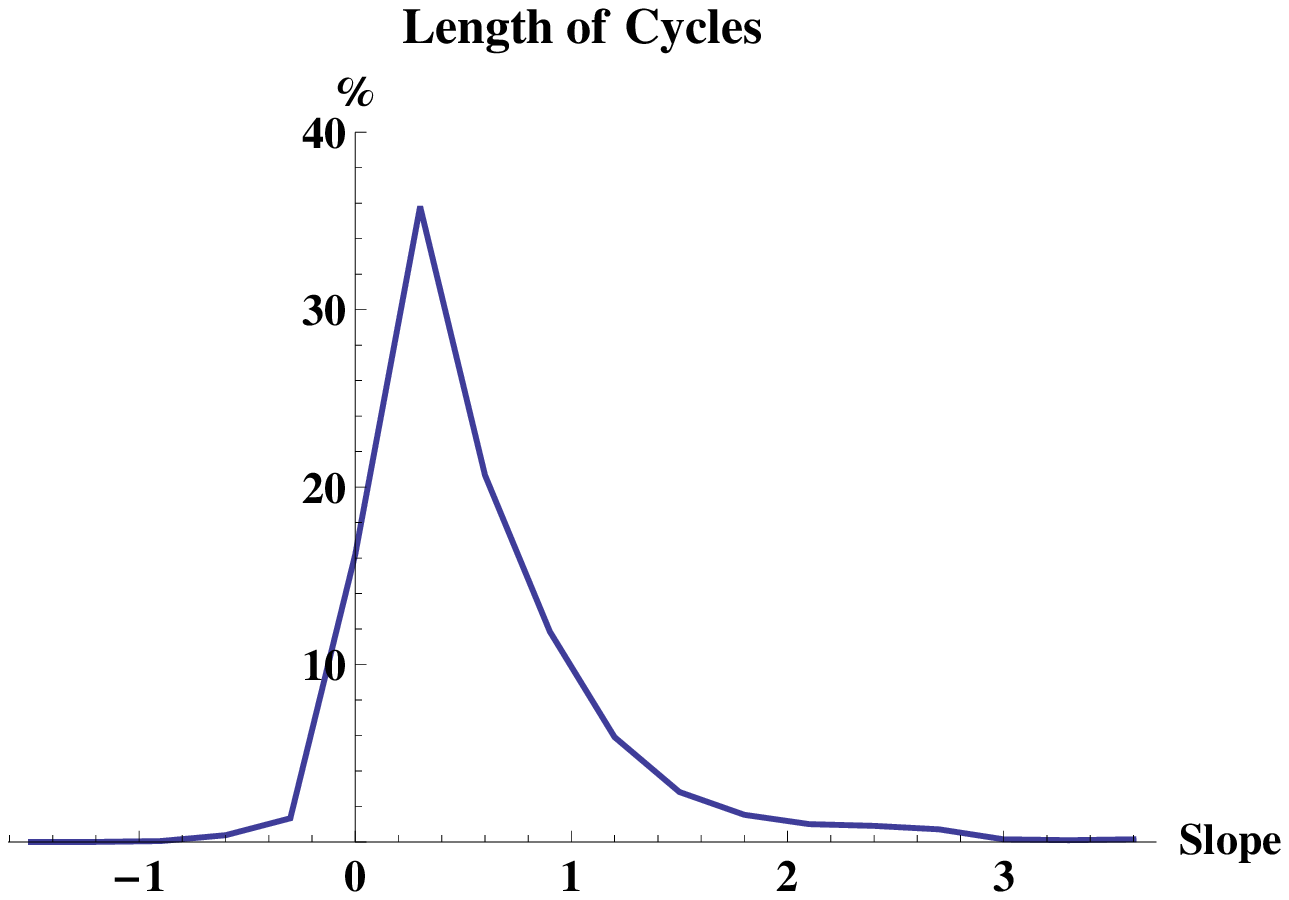}}}
 \raise10pt\hbox{ \framebox{
\includegraphics[width=1.5in]{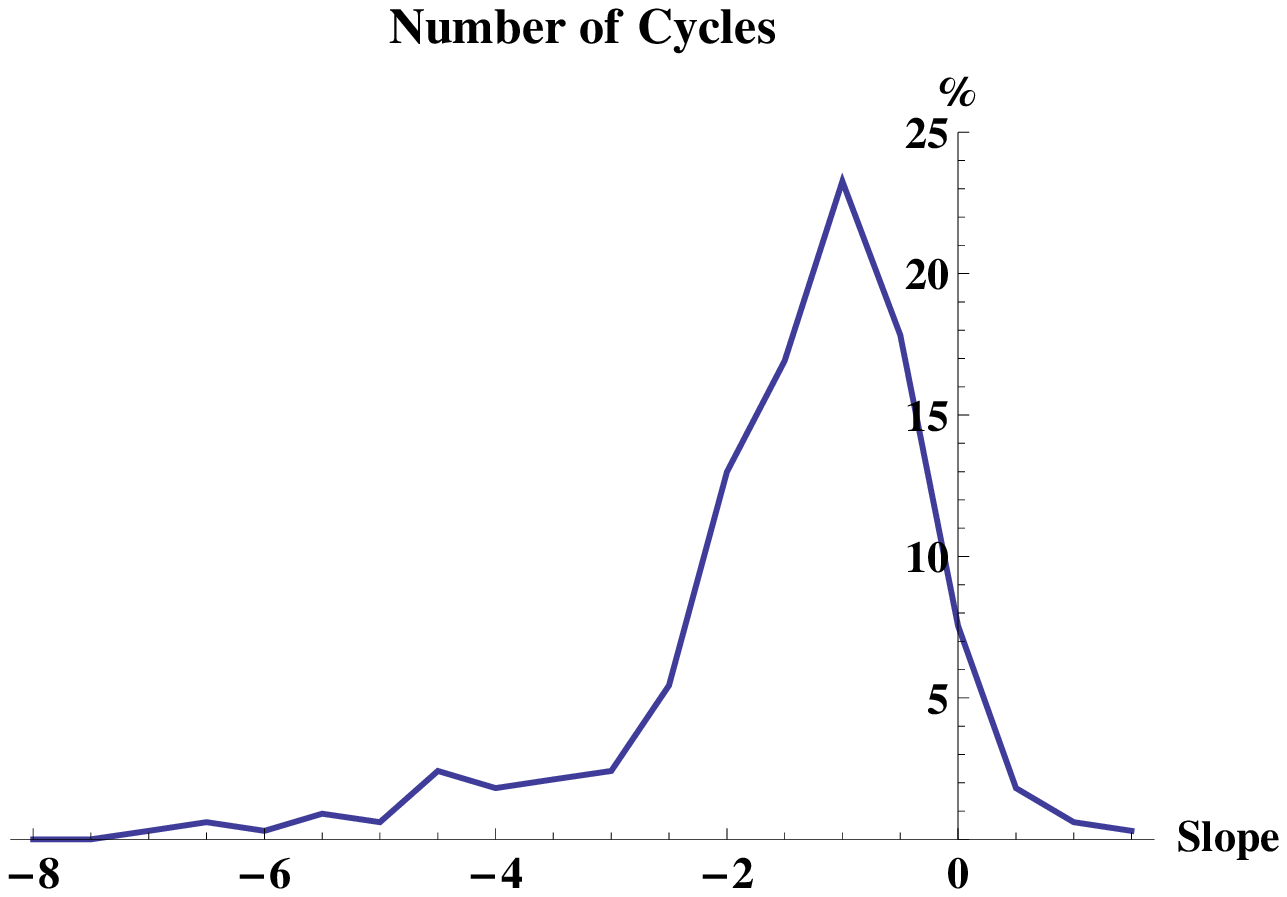}}}
\raise10pt\hbox{ \framebox{
\includegraphics[width=1.5in]{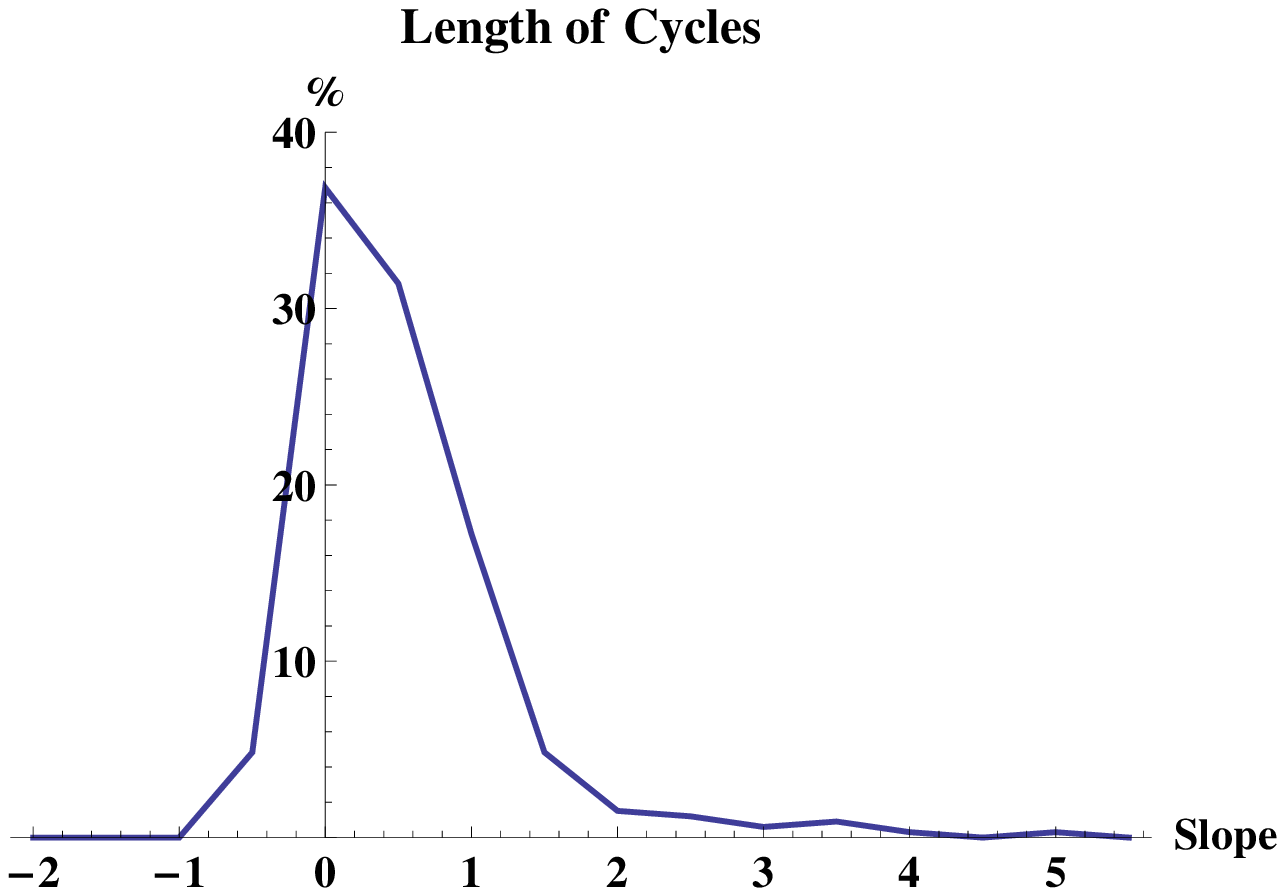}}}
}
\caption{}
\label{15-20-node_results}
\end{figure*}

\subsubsection*{Figure~\ref{15-20-node_results}.}
{Histogram of slopes of best-fit lines to the average number, resp. length, of limit cycles (horizontal axis) vs. percentage of experiments with a given slope (vertical axis).  The left two panels show the results for 15-node networks, the other two panels those for 20-node networks.}


\clearpage
\newpage

\clearpage
\newpage


\clearpage
\newpage
\begin{table}[bpth]
\begin{center}
\caption{}
\label{av_tab}
\begin{tabular}{|l|c|c|}\hline
  d & Av. Num. & Av. Len.\\ \hline
  0 & 3.5 & 1.23 \\ \hline
  1 & 2.80 & 1.25 \\ \hline
  2 & 2.50 & 1.52 \\ \hline
  3 & 1.20 & 3.50 \\ \hline
\end{tabular}
\end{center}
\end{table}

\clearpage
\newpage

\clearpage
\newpage
\begin{table}[bpth]
\begin{center}
\caption{}
\label{total}
\begin{tabular}{|c|c|c|c|} \hline
  n     & Num. of Exp. & Av. Num.   & Av. Len. \\ \hline
  5     & 117000        & 99.75     & 99.83 \\ \hline
  7     & 5000          & 97.82     & 99.92 \\ \hline
  10    & 6000         & 95.70     & 99.58 \\ \hline
  15    & 2921         & 95.72     & 98.25 \\ \hline
  20    & 331          & 90.03     & 94.86 \\ \hline
  100   & 659         & 77.39     & 93.93 \\ \hline
\end{tabular}
\end{center}
\end{table}

\clearpage
\newpage
\begin{table}[bpth]
\begin{center}
\caption{}
\label{tab_5}
\begin{tabular}{|c|c|c|c|c|} \hline
  $D$ & Av. Num. & Av. Len. & Med. Num. & Med. Len. \\ \hline
  $25\%$ & 26 & 114 & 29 & 542 \\ \hline
  $50\%$ & 4 & 16 & 18 & 59 \\ \hline
  $75\%$ & 0 & 1 & 6 &  3  \\ \hline
  $100\%$& 1 & 0 & 2 & 0 \\ \hline
\end{tabular}
\end{center}
\end{table}

\clearpage
\newpage

\clearpage
\newpage

\clearpage
\newpage

\end{document}